\documentclass{PoS}
\usepackage{subfig}

\newcommand{\ie}{{\it i.e.}}

\title{Physics case for a polarised target for AFTER@LHC}

\ShortTitle{A polarised target for AFTER@LHC}

\author{\speaker{J.P.~Lansberg}$^1$, 
M.~Anselmino$^2$, 
R.~Arnaldi$^2$, 
S.J.~Brodsky$^3$, 
V.~Chambert$^1$, 
J.P.~Didelez$^1$, 
M.G~Echevarria$^4$, 
E.G.~Ferreiro$^5$, 
F.~Fleuret$^6$,  
Y.~Gao$^{7}$, 
B.~Genolini$^1$, 
C.~Hadjidakis$^1$, 
I.~H\v{r}ivn\'{a}\v{c}ov\'{a}$^{1}$,
D. Kikola$^{8}$,
A. Kurepin$^{9}$,
C.~Lorc\'e$^{10}$,
L.~Massacrier$^{1,11}$, 
A. Nass$^{12}$, 
C.~Pisano$^{13}$
I.~Schienbein$^{14}$,
M.~Schlegel$^{15}$,
E.~Scomparin$^2$,
J.~Seixas$^{16}$,
A.~Signori$^{17}$,
E.~Steffens$^{18}$,
N.~Topilskaya$^{9}$,
B.~Trzeciak$^{19}$
U.I.~Uggerh\o j$^{20}$, 
A.~Uras$^{21}$,
R.~Ulrich$^{22}$, 
and Z.~Yang$^{6}$.

\\
$^1$ IPNO, Univ. Paris-Sud, CNRS/IN2P3, Universit\'e Paris-Saclay,  Orsay, France\\
$^2$ Dip. di Fisica and INFN Sez. Torino, Via P. Giuria 1, Torino, Italy \\
$^3$ SLAC National\,Accelerator\,Laboratory, Stanford University, Menlo Park, USA\\
$^4$ ECM, Universitat de Barcelona,  Barcelona, Spain\\
$^5$ Dept. de F{\'\i}sica de Part{\'\i}culas, USC, Santiago de Compostella, Spain\\
$^6$ LLR, \'Ecole Polytechnique, CNRS/IN2P3,   Palaiseau, France\\
$^{7}$ CHEP, Department of Engineering Physics, Tsinghua University, Beijing, China\\
$^{8}$ Faculty of Physics, Warsaw University of Technology,  Warsaw, Poland \\
$^{9}$ Institute for Nuclear Research, Russian Academy of Sciences, Moscow, Russia\\
$^{10}$ CPhT, Ecole Polytechnique, CNRS, Universit\'e Paris-Saclay,  Palaiseau, France\\
$^{11}$ LAL, Univ. Paris-Sud, CNRS/IN2P3, Universit\'e Paris-Saclay,  Orsay, France\\
$^{12}$ Institut f\"ur Kernphysik, Forschungszentrum J\"ulich, J\"ulich, Germany \\
$^{13}$ Dipartimento di Fisica, Universita degli Studi di Pavia, Pavia, Italy\\
$^{14}$ LPSC,  Univ. Joseph Fourier, CNRS/IN2P3/INPG,  Grenoble, France\\
$^{15}$ Institute for Theoretical Physics, T\"ubingen U.,  T\"ubingen, Germany \\
$^{16}$ LIP and IST, Lisbon, Portugal \\
$^{17}$ Nikhef and Dept. of Physics and Astronomy, VU  Amsterdam,  Amsterdam, The Netherlands\\
$^{18}$ Physics Institute, Friedrich-Alexander University Erlangen-N\"rnberg, Erlangen, Germany\\
$^{19}$ Institute for Subatomic Physics, Utrecht University, Utrecht, The Netherlands\\
$^{20}$ Department of Physics and Astronomy, University of Aarhus, Denmark\\
$^{21}$ IPNL, Universit\'e Claude Bernard Lyon-I, CNRS/IN2P3, Villeurbanne, France\\
$^{22}$ Institut f\"ur Kernphysik, Karlsruhe Institute of Technology (KIT), Karlsruhe, German
       }

\abstract{We review a number of ideas put forward in favour of the use of 
a polarised target along with the proposed idea  of a fixed-target experiment 
using the LHC beams -- AFTER@LHC. A number of recent studies have shown that 
single transverse-spin asymmetries (STSAs) are large  enough to be precisely 
measured in the region accessible with AFTER@LHC, in particular as 
regards the Drell-Yan process as well as single-pion, isolated-photon and jet production.
 AFTER@LHC with a polarised target would also be the ideal experimental set-up to measure 
the gluon Sivers effect  via a number of original quarkonium STSA studies. We discuss 
first figures-of-merit based on simulations for AFTER@LHC with a polarised target.}

\FullConference{XVIth International Workshop in Polarized Sources, Targets, and Polarimetry, PSTP2015,\\
		14-18 September 2015,\\
		 Bochum, Germany}

\begin{document}

\section{Introduction}

To date, RHIC is the only collider of polarised protons which allows to investigate spin 
physics at ultra relativistic energies, well beyond the mass of most hadronic resonances, 
and in the production of weak bosons. Such measurements were not possible at the Fermilab 
Tevatron nor are at the LHC since their beams are not polarised.
It is however possible to use a polarised target impinged by the LHC beams to perform
 new and unique investigations of single transverse-spin asymmetries (STSAs), $A_N$. 
These are related to the Sivers effect~\cite{Sivers:1989cc} observed since long ago whereby the transverse 
momentum of the partons, $k_T$, is correlated to the nucleon spin. Such a dynamics is 
normally encoded in Transverse-Momentum-Dependent distributions (referred to as TMDs 
in the following) when $k_T$ is explicit or to twist-3 correlators, such as $T_F(x,x)$ 
which is also known as the Efremov-Teryaev-Qiu-Sterman quark-gluon correlator.

In the AFTER@LHC case, owing to the boost between the laboratory and the centre-of-mass 
frames in  the fixed-target mode, measurements of STSAs in the rapidity region covered
 by experiments such as LHCb and ALICE would probe the region of large parton momentum 
fractions in the polarised hadron, $x^\uparrow$~\cite{Brodsky:2012vg,Lansberg:2014myg,Massacrier:2015nsm}. 
This is precisely where the asymmetries are expected to be the largest as recently quantified
for Drell-Yan pair production (DY)~\cite{Liu:2012vn,Anselmino:2015eoa}, isolated 
photon~\cite{Anselmino:2015eoa,Kanazawa:2015fia}, single 
pion~\cite{Anselmino:2015eoa,Kanazawa:2015fia} or jet~\cite{Kanazawa:2015fia} production. 
These STSAs, which are attributed to the dynamics of the quarks inside the polarised hadrons, 
are thus directly (or indirectly, depending on the process) related to the quark Sivers effect. In the DY case, the extracted 
quark Sivers function is predicted~\cite{Collins:2002kn,Brodsky:2002rv} to have an opposite sign 
w.r.t. to the  Semi-Inclusive-Deep-Inelastic-Scattering case. 
The verification of this predicted sign change is the main physics case of the
DY COMPASS run~\cite{Quintans:2011zz} and the experiments 
E1039~\cite{Klein:zoa} and E1027~\cite{Isenhower:2012vh} at Fermilab, 
with a polarised target (resp. beam) to investigate lower (resp. larger) $x^\uparrow$.
AFTER@LHC is an extremely complementary facility to further investigate 
 the quark Sivers effect by measuring DY STSAs~\cite{Liu:2012vn,Anselmino:2015eoa} over
a wide range of $x^\uparrow$ in a single set-up.

As regards the gluon Sivers effect, nearly nothing is known and most of the 
observational constraints are weak, if not model-dependent. The reader is guided 
to~\cite{Boer:2015ika} for a recent overview. Yet, the PHENIX collaboration was 
the first to look at STSAs of gluon sensitive probes with that of $J/\psi$ 
production~\cite{Adare:2010bd}. They found out an $A_N$ compatible with zero but
also allowing for values as large as 20 \%. Let us note that such an $A_N$ is not directly connected
to the gluon Sivers function since $k_T$ is integrated over in $A_N$ and is thus more naturally
connected to twist-3 correlators (see also~\cite{Yuan:2008vn}). More recently COMPASS claimed to have found
a first hint for a nonzero gluonic Sivers effect in di-hadron muoproduction on transversely polarised
protons~\cite{Kurek:2016nqt}. With the large
luminosities offered by AFTER@LHC, similar STSAs for bottomonia, spin-singlet states such as
the $\eta_c$~\cite{Schafer:2013wca} or associated production~\cite{Lansberg:2014myg} 
become accessible paving the way for a convincing program to measure the gluon Sivers 
effect in order to assess the importance of the gluon orbital angular momentum 
in the resolution of the nucleon-spin puzzle.


\section{Two approaches for a fixed-target experiment at the LHC with a polarised target}

We briefly discuss here two promising options allowing the highly energetic LHC beams to 
collide onto a target. Both are currently investigated in the LHC complex. The first 
option relies on the use of a bent crystal which is positioned in the halo of the LHC beam
~\cite{Uggerhoj:2005xz}. 
A few protons (or Pb ions) per bunch per pass can then be channelled along the lattice of
 the crystal. Since it is bent, these particles are deflected by a few mrad w.r.t the beam axis 
along the crystal curvature.  This extraction technique is efficient 
and it allows one to obtain a clean and well collimated beam. A remarkable asset of this 
technique is that it does not affect the LHC performances since, by essence, it only uses  
beam particles which would  anyway be lost for collisions. Such a technique, used in a 
similar manner, is in fact currently investigated as a possible means to improve 
the LHC beam collimation by the (L)UA9 collaboration~\cite{LHCC2011}: two goniometres 
and two bent crystals has 
so far been installed in the LHC beam pipe at IR7 and tested with 
success~\cite{Scandale:2060391}.  This follows other successful data takings for
 protons at  CERN (SPS)~\cite{Arduini:1997kh}, Fermilab~\cite{Asseev:1997yi}, 
Protvino~\cite{Afonin:2012zz} and for lead ions at  CERN SPS~\cite{Scandale:2011za}.

The second option relies on the use of an internal gas target which can be polarised, 
unlike an internal wire target as used at DESY by Hera-B~\cite{Ehret:2000df} and discussed
for the LHC in~\cite{Kurepin:2011zz}. The low density of the gas also allows for a parasitic 
functioning where the beam lifetime is essentially not affected, yet with very large luminosities 
(see below). Such an internal gas-target option is in fact currently used by the LHCb 
collaboration but as a luminosity monitor~\cite{FerroLuzzi:2005em} (SMOG) 
initially designed to monitor the transverse size of the beam.  Even with a limited pressure
(about $10^{-7}$~mbar), it demonstrates the viability of such a solution over extended 
periods of time, without any interferences on the other LHC experiments. LHCb performed 
pilot runs with $p$ and Pb beams 
on a Ne gas target in 2012 and 2013. With the corresponding beam energies, the c.m.s 
energy was $\sqrt{s_{NN}} = 86.6$~GeV for $p$ and 54.5~GeV for Pb. The success of such 
pilot runs motivated longer runs in 2015: $p$ on Ne (12 hours), He (8 hours) and Ar (3 days) 
at 110.4 GeV as well as Pb on Ar (1 week) and $p$ on Ar (a few hours) at 68.6~GeV. No decrease of the LHC performances was 
observed. We stress that the current gas pressure is limited by the pumping system. 
Using a system like the one of the HERMES experiment at DESY-HERA~\cite{Airapetian:2004yf},  
higher pressures can be obtained  with the further possibility of 
polarised gases~\cite{Barschel:2015mka}.

With the bent-crystal option, an average of 15 protons each 25 ns should be extracted. 
This means a flux of 5 $\times$ $10^{8}$ $p^{+}$s$^{-1}$ entirely from the LHC beam 
losses\footnote{We note that contrary to some 
statements made in the 90's, the degradation of the crystal is negligible as  
shown in~\cite{Baurichter:2000wk}. It should be at the level of 6$\%$ per $10^{20}$ 
particles/cm$^{2}$, \ie~1 year of operation. The crystal can simply be moved 
by less than a millimeter or replaced after a year such  that the beam halo hits an
 intact spot.}. In Tab. 1, we gathered the instantaneous and per annum luminosities 
(assuming a run of 10$^7$s for $p^{+}$ over a year)  
for $p^{+}$ on H and D targets. One obtains integrated luminosities 
as large as 20 fb$^{-1}$ with a 1m-long target of liquid hydrogen; such a number is 
comparable to the data sample collected at 7 and 8 TeV at the collider LHC.

With the internal gas-target option, the luminosity is given by  the product
of the particle current $I$ and the areal density of the polarised storage cell target, $\theta$. 
The density is affected by many parameters, the most important being the flux of the polarised source 
injected into the target cell and the geometry of the cell~\cite{Steffens:PSTP2015}. 
For the LHC geometry~\cite{Steffens:PSTP2015} and assuming the flux from the HERMES target source, 
a density of  $2.5 \times 10^{14}$ cm$^{-2}$ for a 1 m long cell for H at 300 K can be expected. 
Such a target density is far below the densities affecting the proton beam life time. 
In the case of the $p^{+}$  beam, $I$ = 3.14 $\times 10^{18}$ $p^{+}$ s$^{-1}$.
The resulting luminosity is shown in Table 1b. 
The storage cell target can be used as well with unpolarised gas of larger atomic masses 
with the LHC lead beams to study heavy-ion collisions at 72 GeV.
The limit of such a solution is essentially set by the number of beam particles ``consumed''
 by the target over a fill or the data  acquisition of the detector.
We have also gathered the instantaneous and yearly luminosities with such a system 
in Tab. 1b.

\begin{table}
\begin{center}
\subfloat[Bent-crystal option]
{\begin{tabular}{c c c c c c c}
  Beam & Target & Thickness & $\rho$ & A & $\cal{L}$ & $\int{\cal{L}}$  \\ 
   &  & (cm) & (g.cm$^{-3}$) &  & ($\mu$b$^{-1}$.s$^{-1}$) & (fb$^{-1}$.y$^{-1})$  \\ \hline
  $p$ & Liquid H & 100 & 0.068 & 1 & 2000 & 20 \\
  $p$ & Liquid D & 100 & 0.16 & 2 & 2400 & 24 \\
\end{tabular}}\\
\subfloat[Polarised internal-gas-target option]{
\begin{tabular}{c c c c c c  c}
  Beam &  Target & $A$ & Areal density ($\theta$)   & $\cal{L}$ & $\int{\cal{L}}$    \\
       &         & &(cm$^{-2}$)    & ($\mu$b$^{-1}$.s$^{-1}$) & (fb$^{-1}$.y$^{-1})$  \\ \hline
  $p$     &   H      & 1 &$2.5\times 10^{14}$    & 900   &  9 \\
  $p$     &   D      & 2 &$3.2\times  10^{14}$ & 1200 & 12
\end{tabular}}
\caption{Expected luminosities (a) with a 7  TeV proton beam extracted by means of bent crystal on conventional targets 
and (b) with an internal gas-target inspired by the HERMES experiment.\vspace*{-0.5cm}}
\end{center}
\end{table}

As what regards the expectations for the target polarisation, the case of the gas target is simple. By reusing a target like that of HERMES, the effective polarisation  of H or D~\cite{Barschel:2015mka} can be as high as 0.8 ($^3$He can also be used).  In the case of an extracted beam, different choices for polarised targets are possible. 
Given the available space in the underground LHC complex, the polarised target system
 used by COMPASS (see~\cite{Doshita:2004ee} and reference therein) may not be optimal 
because of the large size of the cooling system to maintain the frozen spin mode. 
It seems more convenient to opt for the system of E1039 at Fermilab~\cite{Klein:zoa}, that is a 
UVa-type NH$_3$ DNP target with a strong 5T field. Since a complete comparison 
is presented in~\cite{Steffens:PSTP2015}, we refrain from repeating it here. In general, 
 H or D gas targets offer much  better figures-of-merit with no dilution and 
a high polarisation. Such a comparison assumes that the data-taking periods are similar,  
that the effective gas pressure can be similar to that of the HERMES condition (storage cell target) and
that a conventional polarised target  would have a length of about one meter.

\section{Proposed measurements}

\begin{figure}[hbt!]
\begin{center}\small
\subfloat[$A_N^{J/\psi}$]{
\includegraphics[width=0.5\columnwidth]{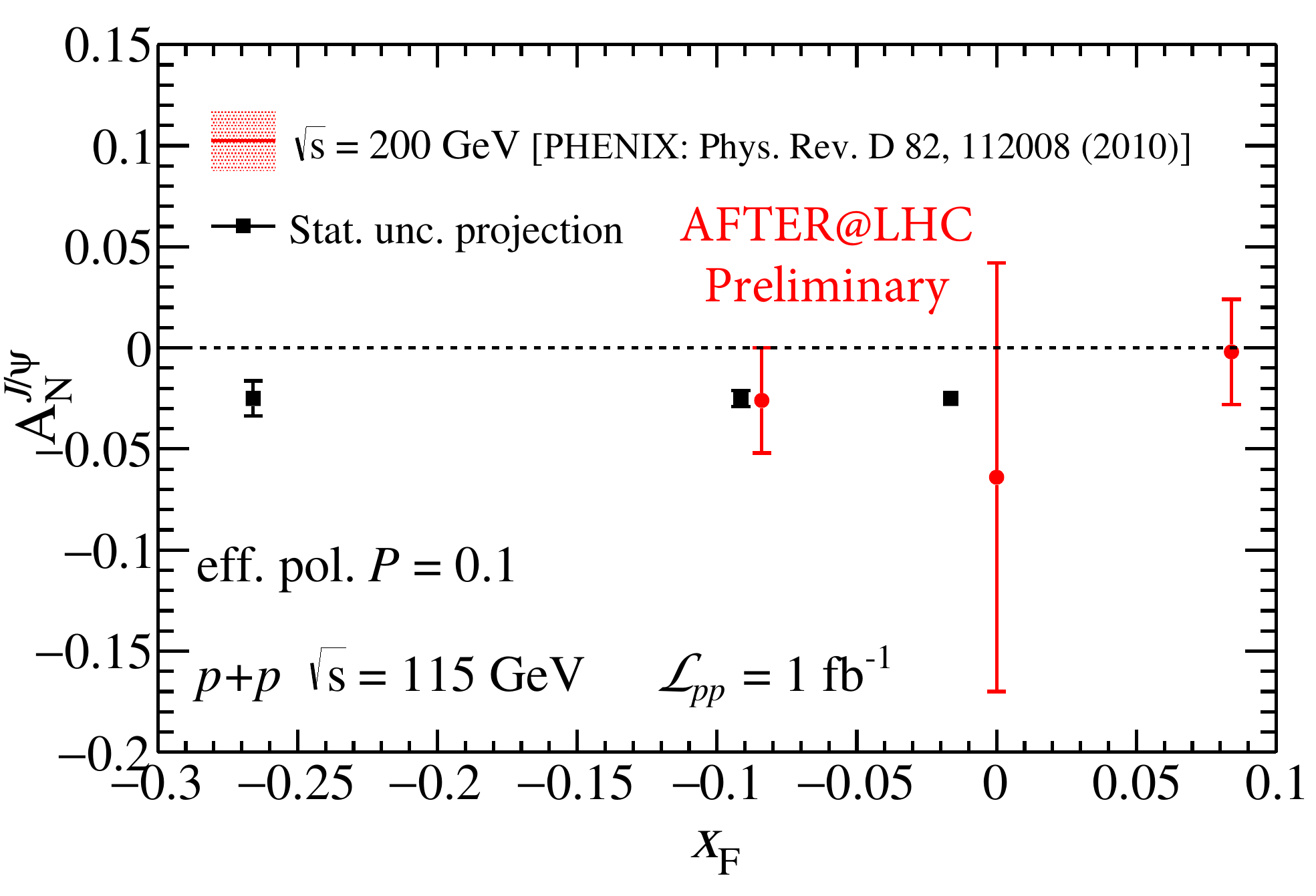}}
\subfloat[$A_N^{\rm DY}$]{
\includegraphics[width=0.5\columnwidth]{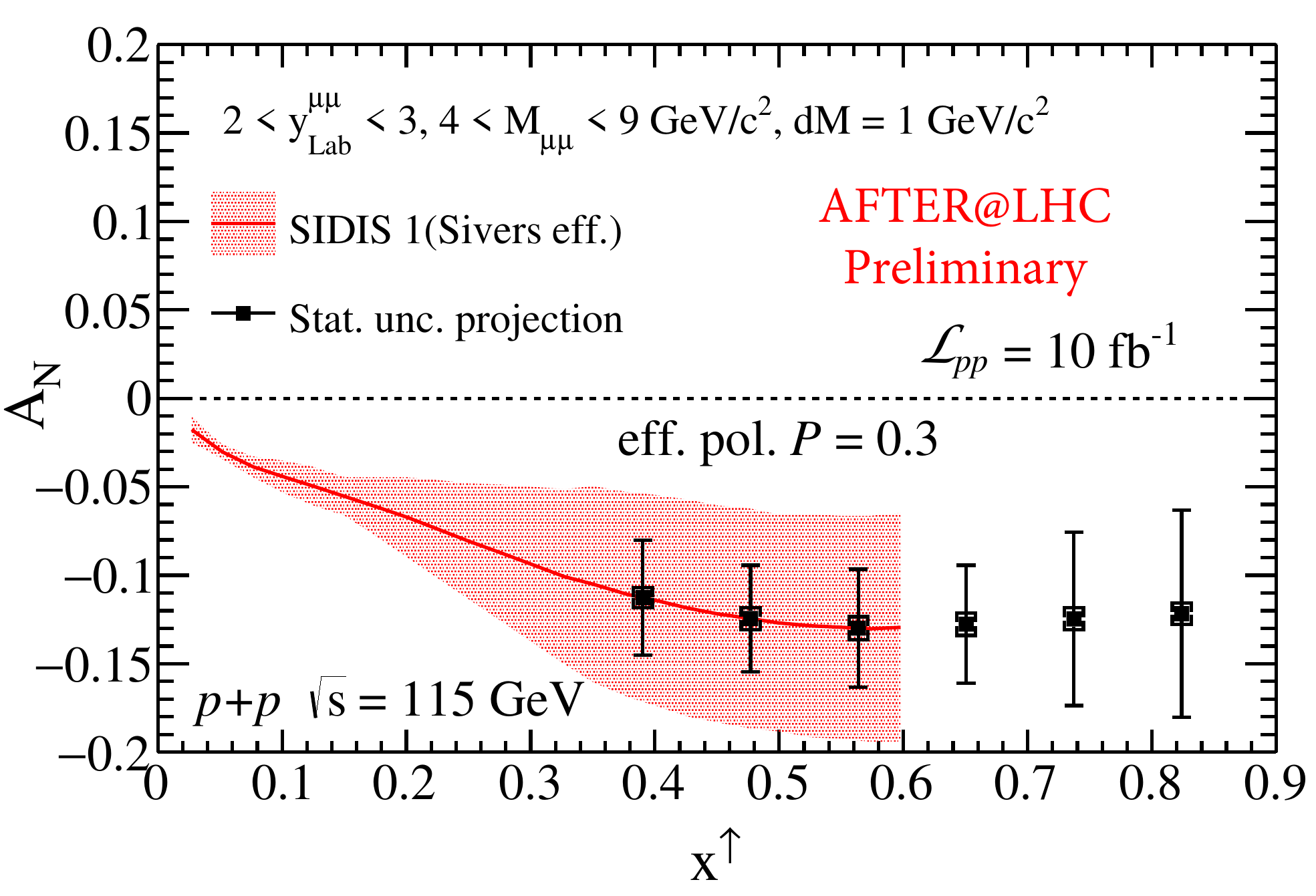}}
\caption{Projected figures-of-merit for STSA of (a) $J/\psi$ and (b) DY production.
The projected uncertainties are based on the indicated effective polarisations and luminosities
using the simulated events of~\cite{Massacrier:2015qba}. In (a), the  3 black points correspond
 to $2<y_{\rm Lab}<3$, $3<y_{\rm Lab}<4$ and $4<y_{\rm Lab}<5$ centred at $A_N^{J/\psi}=0.025$ which is the averaged central values of the PHENIX points. In (b), the theoretical uncertainties in red are 
from~\cite{Anselmino:2015eoa}.\label{fig:An}}
\end{center}
\end{figure}

Recent simulations~\cite{Massacrier:2015qba} have confirmed the very strong case 
for quarkonium physics~\cite{Lansberg:2012kf} with AFTER@LHC. For spin physics, 
measurements of STSAs for single $J/\psi$ or $\Upsilon$ should also be easily accessible
with the set-ups discussed above. Even with less favourable dilution factors or target 
transverse densities (see Fig. \ref{fig:An} (a)), such STSAs should tell us if the Sivers 
effect is suppressed or not in the gluon sector.  More work remains to be done to 
tell down to which $P_T$ other states such as $\eta_c$ or $\chi_c$ could be observed. 
Along the lines of~\cite{Boer:2015ika}, the latter measurements are very important to set
direct constraints on the gluon Sivers function, instead of on twist-3 correlators.
We note that these are also interesting~\cite{Boer:2012bt} to extract gluon TMDs 
of unpolarised proton, similar to the Boer-Mulders functions~\cite{Boer:1997nt},  without a polarised target.

Another very promising observable to unveil the gluon sector of the TMD is certainly the 
production of a pair of $J/\psi$  since both  are produced colourless at low
momenta~\cite{Lansberg:2014swa} and it also remains a gluon-induced process even in the 
far backward domain. The yields for $J/\psi+J/\psi$ computed 
in~\cite{Lansberg:2015lva} are large enough such that one can measure the 
corresponding STSA with sufficient precision and differential in the pair-transverse 
momentum which is necessary 
to access the $k_T$ of the colliding gluons. Such measurements could also be 
very well complemented
 by that of $J/\psi$ or $\Upsilon$ produced back to back to a photon~\cite{Lansberg:2014myg} 
(see also~\cite{Dunnen:2014eta} for a discussion of the interest of such a measurement 
with an unpolarised target) or even di-photon production, as suggested in~\cite{Qiu:2011ai},
 if the background can be dealt with. We therefore plan to carry out 
careful analyses of the projected yields and asymmetries for these reactions.

While analysing the dimuon continuum between the charmonium and bottomonium 
families in~\cite{Massacrier:2015qba},   
we also noticed  that the DY signal is perfectly 
accessible with a tractable combinatorial background which can probably be subtracted 
using the like-sign and/or mixed-event methods. Care should however be taken about 
the heavy-flavour background from charm and beauty. The asymmetries induced by these 
processes can nevertheless  be studied separately by selecting displaced dimuons with a 
vertex detector. In fact, heavy-flavour STSAs
are interesting on their own~\cite{Anselmino:2004nk}. 
In view of this, we expect DY STSAs to be accessible in a very interesting 
kinematical region and to provide unique constraints on the quark Sivers function
especially at large $x$ (large $Q$) where the background is by far the least problematic
(see Fig. \ref{fig:An} (b)). 

As discussed in the introduction, the studies of STSAs of isolated photons, single 
pions as well as jets were nicely motivated in recent 
works~\cite{Kanazawa:2015fia,Anselmino:2004nk}. For these three processes, the yields
are obviously larger than for (open and hidden) heavy-flavour and DY production. 
Since the expected statistical samples are large, the remaining challenge could be to
 gain a sufficient control over the background in the measurement.


\section{Conclusion}

The measurements of STSAs at a fixed-target set-up using the LHC beams
with a detector covering rapidities from roughly 2 to 5 correspond 
to a region where asymmetries are expected to be significant. With both techniques 
which we discussed to study such collisions, namely an extracted beam by means of a bent crystal 
on a polarised target or an internal polarised gas target, the luminosity and
the target polarisation characteristics will be good enough for decisive measurements 
both on quark and gluon sensitive probes. This naturally allows for an 
extended spin-physics program. In particular, the measurements which we mentioned allow for the study of
the physics of initial and final state interactions which opens up 
an entire range of new leading-twist physics and factorisation-violating dynamics for QCD. 
  If a system like the HERMES
target is used, the performances probably exceed those of all the currently running and 
planned  experiments.

\section*{Acknowledgements}
This research [SLAC-PUB-16474] was supported in part by the French P2IO Excellence Lab-
oratory, the French CNRS via the grants PICS-06149 Torino-IPNO, FCPPL-Quarkonium4AFTER \&
D\'efi Inphyniti--Th\'eorie LHC France and by the Department of Energy, contract DE--AC02--76SF00515.

\end{document}